\def\numa{\nu_{max}}
\def\nul{\nu_{L}}
\def\xo{x_{obs}}
\def\jve{J_{-21}(\zo)}
\def\nib{N_{PBH}(z)}
\def\zo{z_{obs}}

\def\oigm{\Omega_{IGM}}
\def\flug{J_{-21}}

\def\lia{Ly_{\alpha}}
\def\oigm{\Omega_{IGM}}

\def\j9{J_{912}}
\def\tgp{\tau_{GP}}

\def\as1{\overline{A_{1}}}

\def\bs1{\overline{B_{1}}}

\def\hab{\hbar}

\def\cosi{\hbar c^{3}}
\def\cosq{\hbar  c^{4}}

\def\part{\partial t}

\def\paro{\partial \omega}

\def\nuso{n_{\gamma PR}}

\def\parnum{\partial n_{\gamma}}
\def\parp{\partial\nuso}

\centerline{\bf THE GUNN PETERSON EFFECT:}
\centerline{\bf A TEST FOR A BLACK HOLES INDUCED}
\centerline{\bf PHOTOIONIZATION OF THE INTERGALACTIC MEDIUM}
\vskip 2cm
\centerline{\bf{Marina Gibilisco}}\vskip 1mm
\centerline{\it Queen Mary and Westfield College}\vskip 1mm
\centerline{\it Astronomy Unit, School of Mathematical Sciences}\vskip 1mm
\centerline{\it Mile End Road, London E1, 4NS;}
\vskip 5mm
\centerline{\it and Dipartimento di Fisica - Universit\'a di Milano}
\centerline{\it Sezione di Astrofisica,}
\centerline{\it Via Celoria, 16, 20133 Milano, Italy.}
\vskip 6mm
\centerline{\bf Abstract:} 
\vskip 4mm
Many experimental evidences indicate the presence of
a ionizing background radiation flux at large redshifts, whose nature 
is doubtful. A lot of informations about 
the characteristics of such a background can be obtained both from the 
study of the Gunn Peterson effect and from the so-called ``proximity effect''.

In some previous works I suggested the possibility that this ionizing flux comes
from the quantum evaporation of primordial black holes (PBHs): here, I 
discuss the constraints that the experimental measurements put upon the 
free parameters of this reionization model and I try to verify its reliability.
In particular, the radiation intensity of the background
at the hydrogen Lyman edge, as inferred from the proximity effect, 
enables me to determine an upper limit to the PBHs average relics mass;
due to our poor knowledge of the ultimate fate of
the evaporating black holes, this limit represents an important theoretical
information.
In the second part of this paper I study 
the absorption of the ionizing background due to $\lia$ clouds: in particular,
I discuss this phenomenon in presence of different absorption levels and 
I calculate the HI Gunn Peterson optical depth $\tau_{GP}(z)$;
from a comparison with the experimental data of Giallongo et al. 
($\tau_{GP,~HI} < 0.02\pm 0.03$) I obtain a constraint on the intergalactic 
medium density parameter, namely $\Omega_{IGM}$ $<0.020$.

A study of the characteristics of the absorbers is also performed:
I determine the hydrogen gas density $n_{H,c}$ and the 
column density $N_{HI}$ for $\lia$ clouds; a satisfactory agreement with the 
available experimental data is obtained in the case of expanding, 
adiabatically cooled clouds. Finally, the same kind of analysis is performed 
for He II: in this case, the theoretical optical depth I obtain is smaller 
than the preliminary experimental lower limit of Jakobsen et al. 
($\tau_{GP}>1.7$).
\vfill\eject
\centerline {\bf 1. INTRODUCTION}
\vskip 7mm
The Gunn Peterson effect [1] is known since 1965 as 
a test of our knowledge of the ionization history of the Universe: 
the presence of an uniform distribution of neutral hydrogen at a redshift $z$ 
should indeed be proved by the presence of an absorption trough in the 
blueside of the $\lia$ emission spectrum of the observed quasars; in fact, 
the $\lia$ resonance absorption line, ($\lambda$=$1216~\AA$), is shifted by 
a factor $(1+z)$ into the visible, blue part of the spectrum.
However, no trough is experimentally observed, in spite of the 
presumable sharpness of this effect;
this fact probably reveals our unsatisfactory knowledge of the 
status of the Universe during the post-recombination era.

On the basis of the pioneering Schmidt's observations [2] of 
the quasar 3C9 at $z=2.016$, Gunn and Peterson [1] obtained 
an integrated $\lia$ optical depth $\tgp \leq  0.5$, 
corresponding to an intergalactic neutral hydrogen number density
$$
n_{HI}(z)~=~\tgp (z)~(1+z)~(1+2q_{0}z)^{1/2}~\Big( {m_{p}\nu_{\alpha}H_{0}
\over \pi e^{2}f}\Big)~=~
$$
$$
~=~{~\tgp (z)~(1+z)~(1+2q_{0}z)^{1/2}\over
4.14\times 10^{10}~h^{-1}};\eqno(1.1)
$$
In eq. (1.1), $H_{0}$ is the present value of the Hubble constant,
equal to $100 ~h~km~s^{-1}$ $Mpc^{-1}$, $0.4\leq h\leq 1$, $q_{0}=1/2$
is the deceleration parameter, $\nu_{\alpha}=2.46\times 10^{15}~sec^{-1}$ is
the HI $\lia$ frequency, $m_{p}$ the proton mass and $f=0.416$ 
is the oscillator strength corresponding to the $\lia$ transition 
[1].
From eq. (1.1), $n_{HI}$ at $z=2$ is equal to 
$6\times 10^{-11}~h~cm^{-3}$; subsequent works [3]
still lowered this limit, that now approximately is 15 times smaller 
than the Gunn Peterson original estimate; for instance, 
Steidel and Sargent found [4]
$n_{HI} (z=0.0)$ $< 9.0\times 10^{-14}~h~cm^{-3}$. 

These values are smaller that the cosmic abundances: 
the inobservance of the $\lia$ absorption trough is probably attributable 
to a high ionization level of the Universe 
rather than to an effective paucity of neutral hydrogen. 

Many causes of such a ionization have been postulated: 
for instance, shock heating phenomena [5], [6], [7],
high mass stars in primordial 
galaxies and unseen quasars, hidden due to an effect of dust obscuration 
by intervening galaxies [8].

All these hypotheses are indeed constrained by the recent
observations of hight redshift quasars ($z >4$); in fact,
the effectiveness of the Gunn Peterson test increases with the redshift, the
optical depth $\tau_{GP}$ being proportional to $(1+z)^{4.5}$.

Apparently, galaxies and unseen quasars cannot be the only photoionizing
sources: in fact, the average ionizing intensity per unit of frequency and 
steradiant they produce is smaller that the value suggested by the so-called
proximity effect; this effect is seen [9], [10] as a decrement in the
counted number of absorbing $\lia$ clouds in the immediate proximity of the 
known quasars: the $\lia$ forest pattern is reduced as a consequence 
of the higher average ionization characterizing these clouds.

The lower limit for the average intensity $\flug$ [9], [10]
at the hydrogen Lyman limit ($\lambda=912~\AA$) is $\flug~\sim 1$
and it is independent on the redshift in the range $1.8 < z < 3.5$;
however, this lower limit for $\flug$ is yet higher than the maximum flux 
coming from the known, counted quasars [10].

\noindent Thus, probably we need some additional radiation sources:
in the following, I would like to discuss a model in which the evaporation of 
primordial black holes produces the high energy photons flux which ionizes 
the IGM at small redshifts.

This model has been investigated in other papers: in particular, I discussed the
effect of an exponential, late and fast reionization of the Universe 
on the polarization of the Cosmic Microwave Background in [11]
while I discussed the details of the reionization processes induced 
by quantum evaporation of PBHs in [12], [13].
Here I will test the
reliability of this model in the light of the Gunn Peterson effect, 
performing a comparison with the available experimental data.
\vskip 5mm
This paper is structured as follows: in Sect. 2 I discuss the main 
characteristics of the evaporating primordial black holes,
particularly their mass evolution in presence of quarks and gluons jets
emission; I also recall the main equations giving the photon emission spectrum:
more details about the time evolution of the ionization
degree $x$ and of the plasma temperature $T_{e}$ can be found in [13].

\noindent In Sect. 3, I discuss the average ionization intensity
$J_{-21}$ coming from the photons emitted by PBHs during their evaporation:
the lower limit $J_{-21}~\sim 1$ suggested by the proximity effect is 
employed to estimate the average PBH relics mass, $\overline M_{rel}$, left 
after the evaporation of these objects.
The agreement between theory and experiment is only possible for 
a PBH average relics mass $\overline{M_{rel}}\sim 10^{-18}~g$: that means
we need a complete evaporation of these primordial objects, 
the most effective reionization being produced at the end of the PBHs life.
A deeper investigation about this point is under study: in fact, 
the quantum gravity effects might change the BH evolution when its mass
lowers under the Planck mass.

In Sect. 3 I also present the main results of this work:
I calculate the HI Gunn Peterson optical depth $\tgp$, both in the case
of a homogeneous IGM and in the more realistic case of an IGM presenting 
some inhomogeneities; as in ref. [8],
a clumping factor $f$ takes into account the presence of moderate 
overdensities in the IGM. The cases of low ($LA$), medium ($MA$) 
and high ($HA$) photoelectric absorption by $\lia$ clouds are all examined for 
different values of the density parameter $\Omega_{IGM}$, namely
$\Omega_{IGM}=0.010, ~0.015, ~0.020$: larger values produce 
a too high $\tgp$ that disagrees with the experimental data concerning 
high redshift ($z\sim 4$) quasars.

In Sect. 4, I discuss in some detail the clumped structures present in 
the IGM in the form of $\lia$ clouds, I determine the efficiency of ionization 
$G_{H}$ and I present my results for the hydrogen number density $n_{H,c}$,
the related column density 
$N_{HI}$ and the density parameter $\Omega_{L\alpha}$.

Two important quantities are involved in this calculation: 
the Doppler shift parameter $b$ and the cloud temperature $T_{c}$;
some evidence of a positive correlation $b-T_{c}$ and $b-N_{HI}$ was suggested
in ref. [14] and it has been related to the presence of 
expanding, adiabatically cooled clouds in [15];
my analysis seems to confirm such a possibility.

In Sect. 5, I perform the same calculation in the case of ionized helium and
I discuss the relevance of the Gunn-Peterson effect for He II;
the intergalactic helium should be present in the form of singly 
ionized He II rather than in the neutral form, proving again the
high ionization status of the IGM at high redshift.
Jakobsen et al. [16] recently claimed the observation of the
He II $304~\AA$ Gunn-Peterson absorption from diffuse IGM at $z\sim 3.2$,
with a total optical depth $\tau_{GP}$(HeII) $>1.7$; the theoretical 
value I calculated is smaller than this preliminary limit, namely equal 
to 0.76.

Finally, in Sect. 6 I summarize the results obtained by using this model of 
PBHs-induced photoionization of the IGM and I present my conclusions.
\vskip 7mm
\centerline {\bf 2. THE QUANTUM EVAPORATION OF}
\centerline {\bf PRIMORDIAL BLACK HOLES.}
\vskip 7mm
The primordial black holes are very interesting structures, 
from a theoretical point of view; however, the processes 
that caused their formation and, in general, their overall 
evolution are poorly known; that despite of the relevant effects their 
quantum evaporation may have on the present status of the Universe. 

Following a widely accepted idea, the black holes formation 
should characterize the early instants of the Universe 
after the Big Bang; however, the nature of the phenomena acting to create 
such structures is not clear and many theories have been advanced [17].

The creation of a black hole is induced by the contraction 
of a mass to a size less than its gravitational radius; 
large mass stars at the end of their evolution may typically represent
some good candidates to form BHs.
Indeed, a gravitational contraction of such a size is quite unlikely for 
small stars and therefore the formation of black holes having a small mass 
($M\leq 10^{17}~g$) is only possible in presence of a huge compression, 
i.e. at the early stages of the cosmological expansion. 

These small mass, primordial black holes present 
an intense Hawking's quantum evaporation: in fact, the blackbody
temperature of the emission is inversely proportional to their mass 
[18]:
$$
kT~=~{\hbar  c^{3}\over 8\pi GM}~\sim~1.06~\Big[~{M\over 10^{13}~g}~\Big]^
{-1}~GeV.\eqno(2.1)
$$
The nature of the emitted particles clearly depends on the blackbody 
temperature: the mass loss makes 
the BH hotter and hotter, thus enabling the emission of more and more 
massive particles. 
BHs having a mass larger than $10^{17}~g$ can only emit massless particles
[19], [20], [21];
however, when the BH mass falls below $10^{14}~g$, hadrons 
emission becomes possible: for temperatures above the QCD confinement 
scale $\Lambda_{QCD}$, a relevant emission of quarks and gluons jets
does start; thus, the resulting spectrum is no more a blackbody one
[20], [21], [22].

As I discussed in [12], [13],
the photons emitted by PBHs during their quantum
evaporation may be one of the causes of the reionization of the Universe:
indeed, they seem very efficient to produce a late and sudden
(nearly exponential) rise of the ionization degree for a reionization 
redshift $z_{R} \leq 30$, while for $ 30 < z_{R} \leq 60$ they are still
able to cause a partial reionization.

Note that the plasma heating due to the photons / electrons 
interactions is not strong enough to induce a relevant distortion of the
CBR spectrum: that is in agreement with the recent FIRAS upper limit on the 
comptonization parameter, $y_{c}<2.5 \times 10^{-5}$ [23].

The most significant particle emission should happen at the last 
stages of the life of a BH; thus I choose a reionization time which 
corresponds to the evaporation time for an object having a mass 
near the critical one ($M_{c}\sim 4.4\times 10^{14}~h^{-0.3}~g$ is the 
mass of a PBH that survives till the present epoch).
\vskip 4mm
In the following, I will briefly recall the fundamental equations describing 
the quantum evaporation of a PBH: more details can be found in 
[12], [13] and references therein.

The initial mass $M_{i}$ of a PBH is connected to its lifetime by 
the following general formula [22]:
$$
t_{evap}~\sim~1.19\times 10^{3}~{G^{2}M_{i}^{3}\over \cosq ~f(M_{i})}~\sim~
6.24\times 10^{-27}~f(M_{i})^{-1}~M_{i}^{3}~sec;\eqno(2.2)
$$
the function $f(M)$ in eq. (2.2) contains the contributions of 
the different species of particles, it is normalized to the unit for 
very massive ($M \geq 10^{17}~g$) BHs and reads as follows:
$$
f(M)~=~1.569~+~0.569~\Bigg[~exp\Big[~{-M\over {4.53\cdot 10^{14}}}~\Big]_{\mu}~
+~6~exp\Big[~{-M\over {1.60\cdot 10^{14}}}~\Big]_{u,d}~+~
$$
$$
+~3~exp\Big[~{-M\over {9.60\cdot 10^{13}}}~\Big]_{s}
~+~3~exp\Big[~{-M\over {2.56\cdot 10^{13}}}~\Big]_{c}~
+~exp\Big[~{-M\over {2.68\cdot 10^{13}}}~\Big]_{\tau}~+
$$
$$
~+~3~exp\Big[~{-M\over {9.07\cdot 10^{12}}}~\Big]_{b}
~+~3~exp \Big [~{-M\over {0.48\cdot 10^{12}}}~\Big ]_{t}~ \Bigg]~+~
$$
$$
+~0.963\Big[~exp\Big[~{-M\over {1.10\cdot 10^{14}}}~\Big]_{gluons} 
\Big ].\eqno(2.3)
$$
In eq. (2.3) the first addendum in the right-hand side expresses the 
contribution of electrons, positrons, photons and neutrinos;
heavier particles are considered in the remaining terms, following their 
relative importance and with a factor 3 taking into account the color charge
for the quarks; the denominators in the exponential terms are defined as 
the product $\beta_{sj}M_{j}$, where $M_{j}$ is the mass of a black hole 
whose temperature is equal to the rest mass $\mu_{j}$ of the $j$ specie and 
$\beta_{sj}$ is a spin-dependent factor defined [22] in such a way the energy 
of a BH having $M=\beta_{sj}M_{j}$ has a peak at $\mu_{j}$.
               
When the BH mass falls below $10^{14}~g$ and the temperature 
$T$ becomes larger than the confinement scale $\Lambda_{QCD}$, it is no longer 
possible to neglect the quarks and gluons emission in the calculation
of the BH mass evolution: following [20], [21], [22],
one should write this evolution as
$$
{dM\over dt}~=~-\sum_{j} {1\over 2\pi\hab}~\int \Gamma_{j}~\Bigg[~exp
~\Big[~{8\pi GQM\over \cosi}~\Big]~-~(-1)^{2s_{j}}~\Bigg]^{-1}\times
{Q~dQ\over c^{2}};\eqno(2.4)
$$
in eq. (2.4), $\Gamma_{j}$ is the 
absorption probability for the $j$ particle having a spin $s_{j}$ [24] and 
one sums on all the emitted species [22]; this equation
means the emission of a parent particle $j$ with total energy $Q$ 
decreases the BH mass by $Q/c^{2}$. After the integration over the energy 
$Q$, eq. (2.4) can be rewritten as [22]:
$$
{dM\over dt}~=~-5.34\times 10^{25}~f(M)~M^{-2}~g~sec^{-1}.\eqno(2.5)
$$
Now, the Hawking emission rate of particles having an 
energy in the range $(E, E+dE)$ from a black hole having an angular velocity 
$\omega$, an electric potential $\phi$ and a surface gravity $\kappa$ is 
[18]:
$$
{dN\over dt}~=~{\Gamma ~dE\over 2\pi\hbar }~\Bigg[~exp~\Bigg(~
{{E-n\hbar  \omega-e\phi}\over{\hbar \kappa/2\pi c}}~\Bigg)~\pm~1~\Bigg]^{-1},
\eqno(2.6)
$$
where the signs $\pm$ respectively refer to fermions and bosons
and $\Gamma$ is the absorption probability of the emitted species.
In the case of $photon$ emission, 
$\Gamma$ reads [24]:
$$
\Gamma_{s=1}~=~{4A\over 9\pi}~\Big({M\over M_{PL}}\Big)^{2}~
\Big({\omega\over \omega_{PL}}\Big)^{4};\eqno(2.7)
$$
in eq. (2.7) $A$ is the surface area of the BH and the Planck mass 
and energy assure we are working with dimensionless quantities, 
as in [24].

Here, I neglect the charge and the angular momentum of PBHs,
a quite reasonable and simplifying assumption due to the fact 
their loss via the quantum evaporation happens on a time scale shorter than the 
one characterizing the mass evaporation [25].

As pointed out in [26] in the emission spectrum we need a 
fragmentation function in order to take into account the production of 
quarks and gluons jets:
$$
{dN_{x}\over dtdE}~=~\sum_{j}~\int^{+\infty}_{0}~{
\Gamma_{j}(Q,T) \over 2\pi\hbar }~\Big(~exp{Q\over T}\pm1~\Big)^{-1}~
{dg_{jx}(Q,E)\over dE}~dQ;\eqno(2.8)
$$
here $x$ and $j$ respectively label the final and the directly emitted 
particles while the last factor, containing the 
fragmentation function $g_{jx}$, expresses
the number of particles with energy in the range $(E, E+dE)$ coming
from a jet having an energy equal to $Q$ [26]:
$$
{dg_{jx}(Q,E)\over dE}~=~{1\over E}~\Bigg(~1-{E\over Q}~\Bigg)^{2m-1}~
\theta(E-km_{h}c^{2});\eqno(2.9)
$$
in eq. (2.9) $m_{h}$ is the hadron mass, $k$ is a constant $\sim O(1)$
and $m$ is an index equal to 1 for mesons and 2 for baryons. 

After selecting the dominant contribution to the integral over $Q$ and 
summing over the final states, eq. (2.8) becomes [26]:
$$
{dN\over dtdE}~\sim~E^{2}~exp \Big({-E\over T}\Big) ~~~~~~~~~~~~~for
~~E >> T~~~~~~ ~Q\sim E, \eqno(2.10a)           
$$
$$
~~~~~~~{dN\over dtdE}~~\sim~ E^{-1}~~~~~~~~~~~~~~~~~~~~ for ~~T\sim E >> m_{h}~~
~~~~Q\sim T,\eqno(2.10b)
$$
$$
~~~~~~~~~{dN\over dtdE}~\sim~{dg \over dE}~~~~~~~~~~~~~~~~~~~~~for 
~~E\sim m_{h} << T~~~~~~
Q\sim m_{h};\eqno(2.10c)
$$
eqs. (2.10a), (2.10b) and (2.10c) respectively hold for the dominant value 
of $Q$ written on the right.
\vskip 7mm
\centerline {\bf 3 THE IONIZING PHOTON FLUX}
\vskip 7mm
The possible origin of the ionizing flux 
has been studied in many papers [8], [27], [28], [29], [30], [31].
Generally, quasars are coinsidered the best 
candidate sources for the photoionization of the Universe;
however, a background generated by quasars only 
is not consistent with the predictions of the proximity effect and 
with the data concerning the evolution of $\lia$ clouds. 
For instance, in ref. [31] the Authors claimed that these quasars  
should ionize the IGM too late and produce a too large Gunn Peterson optical 
depth (note however that their conclusions have often been questioned).

In the following, I will discuss the mean specific intensity of the 
radiation field coming from the evaporation of PBHs as a function of the final 
mass of their relics; then, I will calculate the Gunn Peterson optical 
depth associated with the resonant HI $\lia$ absorption.
\vskip 7mm
\centerline {\bf 3.1 THE RADIATION INTENSITY}
\centerline {\bf AT THE HYDROGEN LYMAN EDGE}
\vskip 7mm
At the hydrogen Lyman edge, $\nu_{L}=c/912~\AA~=~3.29\times 10^{15}~sec^{-1}$, 
and for an observer at a 
redshift $z_{obs}$,
the mean intensity of the photons flux in $erg~cm^{-3}~sec^{-1}~Hz^{-1}~
sr^{-1}$ is [32]:
$$
J_{912}(z_{obs})~=~{c\over 4\pi H_{0}}~\int^{zmax}_{zobs}~{(1+\zo)^{3}\over
(1+z)^3}~{\epsilon_{T}(\nu,z)~exp~[-\tau_{eff}(912,\zo ,z)]\over (1+z)^{2}~
(1+2q_{0}z)^{0.5}}~dz;\eqno(3.1.1)
$$
in eq. (3.1.1) $\epsilon_{T}(\nu,z)$ is the total emissivity of a 
primordial population of BHs, calculated at a frequency $\nu=\nu_{L}~(1+z)~
(1+\zo)^{-1}$ and at a redshift $z$, expressed in $erg~cm^{-3}~$ 
$sec^{-1}~Hz^{-1}$; $\tau_{eff}(912, z_{obs}, z)$ is an
effective optical depth that, as I will discuss below, takes into account 
the probability that these photons are absorbed.

The emissivity $\varepsilon$ can be obtained by calculating the 
total photon number density from eq. (2.10c).
In our case, the condition $E << T$ holds: we are effectively 
interested in the last stages of the evaporation, when the BHs temperature
is very high; moreover, the processes that are relevant for this study 
(the ionization and the recombination) dominate [20], [21]
for an energy $E\leq 14~KeV$.

For a jet fragmentation function given by eq. (2.9), the photon density
in the unit time and energy can be written as follows:
$$
{\parnum\over \paro\part}|_{Tot}~=~{\parnum\over \paro\part}|_{Mes}~+~
{\parnum\over \paro\part}|_{Bar},\eqno(3.1.2)
$$
where the mesons and baryons contributions have been singled out.

For $Q=m_{hadr}\sim 300~MeV$, eq. (3.1.2) reads 
$$
{\parnum\over \paro\part}|_{tot}~=~{1\over \omega}~\Big(1-{\omega\over Q}
\Big)~+~
{1\over \omega}~\Big(1-{\omega\over Q}\Big)^{3},\eqno(3.1.3)
$$
the dominating contribution being the mesonic one; 
after reducing to the proper dimensions, I obtain the volume emissivity 
of a $single$ evaporating PBH:
$$
{\parp\over\partial\tilde t\partial\tilde\omega}~=~\epsilon (\omega)~=~
{3.20\times 10^{-15}\over \omega~(GeV)}~erg~cm^{-3}~sec^{-1}~Hz^{-1}.
\eqno(3.1.4)
$$
After considering the frequency shift
$\nu=\nu_{L}~(1+z)~(1+\zo)^{-1}$, eq. (3.1.4) becomes:
$$
\epsilon(\nu,z)~=~{3.36\times 10^{-40}\over \nu_{L}}~{(1+\zo)\over (1+z)}~
erg~cm^{-3}~sec^{-1}~Hz^{-1}.\eqno(3.1.5)
$$
Finally, the total emissivity is:
$$
\epsilon_{T}(\nu,z)~=~N_{PBH}(z)~\epsilon (\nu,z),\eqno(3.1.6)
$$
where the parameter $\nib$ represents the population of PBHs at a 
redshift $z$.

Its value can roughly be estimated as follows: I write the
PBHs density as
$$
\rho (z)~\sim~{\overline{M_{rel}(z)}~\nib\over R^{3}(z)},\eqno(3.1.7)
$$
where $\overline M_{rel}(z)$ is the average mass of the PBHs relics at 
a redshift $z$. Then, the scale factor evolves as in the radiation dominated 
epoch:
$$
R(t)~\sim~R_{0}~(t/t_{0})^{1/2};\eqno(3.1.8)
$$
($R_{0}=1.25\times 10^{28}~cm~=1.4\times 10^{10}~lyr$ in a standard
cosmological model, see ref. [33]).

Now, I assume the PBHs density evolution is approximately described by a 
power law with index $2/3$; the formation time coincides with the Big Bang
and the initial density is $\rho_{i}\sim 4.28\times 10^{24}~g~cm^{-3}$,
[13] as one infers on the basis of the present experimental 
limits on $\Omega_{PBH}$ ($\Omega_{PBH} < (7.6\pm 2.6)\times 10^{-9}~
h^{(-1.95\pm 0.15)}$) [20], [21].
Then, one finally gets [13]:
$$
\nib~=~ R_{0}~\rho_{i}~\big({t_{i}\over t_{0}}\big)^{2/3}~(1+z)^{1/4}~{1\over 
\overline M_{rel}(z)},\eqno(3.1.9)
$$
i.e.:
$$
\nib~=~9.67\times 10^{-10}~{(1+z)^{1/4}\over \overline M_{rel}(z)}~cm^{-2}.
\eqno(3.1.10)
$$
The most difficult parameter to estimate in eq. (3.1.10) is the 
average relics mass $\overline M_{rel}$ at the reionization epoch 
($z\sim 20\div 30$); 

This parameter can be evaluated by searching a basic agreement
with the data coming from the proximity effect and
concerning the average intensity of the ionizing background;
an agreement is only possible for $\overline M_{rel} \sim 10^{-18}~g$, i.e. 
for black holes totally evaporated or may be for relics well lighter than the 
Planck mass.
However, for these objects the quantum gravity effects might be very relevant,
thus a deeper theoretical analysis about this point is under 
study. 
\vskip 6mm
Turning back to the effective optical depth, it can be written as [34]:
$$
\tau_{eff}(912, \zo, z)~=~\int^{z}_{\zo}~\int^{\infty}_{0}~
{\partial^{2} N\over \partial N_{HI}\partial z^{'}}~\times
~[1~-~exp (-N_{HI}\sigma_{\nu '})]~dN_{HI}~dz^{'};\eqno(3.1.11)
$$
in eq. (3.1.11) $\sigma_{\nu '}$ is the hydrogen photoionization cross
section, $N_{HI}$ is the hydrogen column density of the 
absorber and the derivative 
$dN / dz$ represents the column density distribution, depending on 
the assumed model of attenuation [32]: in fact,
we can have various level of absorption and 
the approximate integration of eq. (3.1.11),
respectively in the cases of low, medium and high absorption, gives [32]:
$$
\tau^{LA}_{eff}(912, \zo, z)~\simeq~\Big[~0.0118 ~\xo^{3}~(x^{0.4}-\xo^{0.4})~+~
2.35~\xo^{1.5}~ln\Big(~{x\over\xo}~\Big)~-
$$
$$
~-0.78 ~\xo^{3}~(\xo^{-1.5}-
x^{-1.5})~-~0.003~(x^{1.5}-\xo^{1.5})~\Big],\eqno(3.1.12)
$$
$$
\tau^{MA}_{eff}(912, \zo, z)~\simeq~\Big[~0.244 ~\xo^{3}~(x^{0.4}-\xo^{0.4})~+~
2.35~\xo^{1.5}~ln\Big(~{x\over\xo}~\Big)~-
$$
$$
~-0.78~ \xo^{3}~(\xo^{-1.5}-
x^{-1.5})~-~0.003~(x^{1.5}-\xo^{1.5})~\Big],\eqno(3.1.13)
$$
$$
\tau^{HA}_{eff}(912, \zo, z)~\simeq~\Big[~0.097~ \xo^{1.56}~
(x^{1.84}-\xo^{1.84})~-~
0.0068~\xo^{3}~(x^{0.4}-\xo^{0.4})~-~
$$
$$
-~8.06\times 10^{-5}~(x^{3.4}-\xo^{3.4})~\Big],\eqno(3.1.14)
$$
where $\xo=1+\zo$ and $x=1+z$.

The effective nature of the absorbers is still unclear: 
probably, they are large clouds (up to $1~Mpc$) containing 
a significant baryonic fraction [35];
many recent analyses identify such absorbers with evolved densities 
fluctuations in the intergalactic medium, produced by large scale flows and
inhomogeneities of the dark matter component of the Universe [36];
anyway, these absorbers both severely attenuate the flux coming from
the ionization sources and generally delay the growth of HII regions [32].
\vskip 5mm
By using eqs. (3.1.5), (3.1.6) and (3.1.10) in eq. (3.1.1), 
I obtain:
$$
J_{912}(z_{obs})~=~{c\over 4\pi H_{0}}~\int^{zmax}_{zobs}~{(1+\zo)^{4}\over
(1+z)^{25/4}}~exp~[-\tau_{eff}(912,\zo ,z)]\cdot
$$
$$
\cdot~{9.67\times 10^{-10}~\over \overline M_{rel}}~{3.36\times 
10^{-40}\over \nu_{L}}
~dz.\eqno(3.1.15)
$$
After recalling the definition of the reduced intensity
$$
J_{-21}(\zo)~=~J_{912}(z_{obs})/10^{-21},\eqno(3.1.16)
$$
I plot in fig. 1 the behaviour of $\jve$ vs $\zo$ obtained from eq. (3.1.15), 
in the cases of $low$ (LA), $medium$ (MA) and $high$ (HA)
absorption; I take $\overline M_{rel} 
=\overline M_{rel} (z_{reion})~\sim 10^{-18}~g$.

\vskip 6mm

The Gunn-Peterson optical depth associated with the resonant 
$\lia$ absorption can be written in the following form [1]:
$$
\tgp (\zo)~=~\Big(~{\pi e^{2} f_{\alpha}\over m_{e}\nu_{\alpha} H_{0}}~\Big)
{n_{HI,d}(z)\over (1+z)(1+2q_{0}z)^{1/2}},\eqno(3.1.17)
$$
where $n_{HI,d}$ is the HI density of the IGM 
diffuse component and $f_{\alpha}$
is the oscillation strength of the $\lia$ transition.

Rewriting eq. (3.1.17) as a function of the intensity $\jve$
one obtains [32]:
$$
\tgp (\zo)~=~3~T_{4}^{-0.75}~(\Omega_{IGM}~h^{2})^{2}~(3~+~\alpha)~(1+z)^{4.5}
~f(z)~(\jve) ^{-1};\eqno(3.1.18)
$$
eq. (3.1.18) is evaluated by inserting the temperature 
$T_{4}=T/10^{4}~K^{\circ}$ of the ionized intercloud medium, as it results
from the PBHs induced photoionization processes [12], [13].
A plot of $T_{4}$ (in $K^{\circ})$ as a function of the redshift $z$ 
is shown in fig. 2; in eq. (3.1.18) $\alpha$ is the 
power spectrum index of the metagalactic flux at high 
$z$ ($J\propto \nu^{-\alpha}$);
in our case
$$
J_{-21} (\zo, \nu)~=~\Big( {\nu_{L}\over \nu}\Big)~\jve,\eqno(3.1.19)
$$
thus $\alpha=1$.

If compared to the original equation of Gunn and Peterson,
Eq. (3.1.18) contains an additional factor, namely the clumping 
factor $f(z)$: the presence of inhomogeneities in the IGM
may influence in a relevant way the processes of absorption 
[5], [6], [7], [8], [37] and, indeed, the observation of $\lia$ clouds is 
a clear proof of the existence of remarkable overdensities. 

The function $f$ describes moderate overdensities 
($1 < \rho/\overline {\rho} \leq 10$) and it is defined as follows [8]:
$$
{< \rho ^{2} >\over <\rho >^{2}}~\equiv~f,\eqno(3.1.20)
$$
while larger inhomogeneities are directly identified with $\lia$ clouds.

The values of $f$ at some different redshifts are listed in tab. 1;
these values have been obtained by Carlberg and Couchman [38] through 
a numerical simulation.

Another important parameter entering in eq. (3.1.18) is the density 
parameter for the intergalactic medium, $\Omega_{IGM}$: here I tested 
3 values, namely $\oigm$ $=0.010, ~0.015,~0.020$, for all the 
cases of low, medium and high absorption and both for an idealized,
perfectly homogeneous IGM ($f(z)=1$) and in presence of moderate
inhomogeneities ($f(z)$ as in tab. 1). Figs. 3a, 3b, 4a, 4b, 5a, 5b 
show the behaviour of $\tgp$ vs $\zo$ for all these cases; in tab. 
2a, 2b, 3a, 3b, 4a, 4b I listed the values of $\tgp$ calculated for 
two reference redshifts, $\zo=3,~\zo=4.3$, 
for which some experimental data are known; tab. 5 finally gives
a picture of the present experimental $\tgp$ measurements, that also
includes the recent data of Giallongo et al. [39].

Finally, in figs. 6a, 6b and 6c I resume the behaviour of $\tau_{GP}(z)$
for the different values of the density parameter $\Omega_{IGM}$ and for 
various levels of absorption.

\vskip 7mm
\centerline {\bf 3.2 DISCUSSION OF THE RESULTS}
\vskip 7mm
As I told, the average photon intensity $\flug$ determined from the 
proximity effect is employed to constrain the final average mass
of the PBHs relics, that should be nearly totally evaporated at the
time of the reionization.

The calculation of the Gunn Peterson optical depth
is strongly model-dependent: in fact, 
eq. (3.1.18) contains both the temperature of the ionized IGM 
and the power spectrum index $\alpha$ of the metagalactic flux at high $z$,
i.e. two peculiar predictions of the model.

Looking firstly at fig. 1, one can remark that the radiation intensity 
evolution 
depends on the behaviour of the effective optical depth, eq. (3.1.11); the same
behaviour is obtained for $\flug$ in the cases of low and medium 
absorption, only rescaled by a factor weighting the first cubic term 
in eqs. (3.1.12) and (3.1.13). On the contrary, 
in the case of a high absorption level, eq. (3.1.14) for $\tau_{eff}$
produces a steeper curve.

The behaviour of $\flug$ vs. $z$ differs in a sensitive way 
from the one predicted, for instance, in [8],
where the sources of the ionization one assumes are quasars
and high-mass stars in primordial galaxies. At early 
epochs, the intergalactic gas is neutral and the Universe is opaque 
to the radiation, while the volume emissivity of photons, $\varepsilon 
(\nu,z)$, is not so high to produce a relevant ionization; thus, 
the intensity $\flug$ in fig. 2 goes to zero at high redshifts.

In ref. [8] the rise of the intensity $J_{-21}$ begins at $z\sim 5$ and stops 
at $z\sim 2.5$, due to fact that, at this redshift, the quasars 
start to decline; on the contrary, the behaviour of the intensity
shown in fig. 1 is nearly exponential, due to the peculiar emission process
I considered.

Looking at tabs. 2a, 2b, 3a, 3b, 4a and 4b, 
together with the experimental 
available data for the Gunn Peterson 
optical depth shown in tab. 5, we can observe that:

\item{a)} the experimental data generally refer to different values of the 
redshift $z$ and therefore, in order to allow a comparison, 
I considered two reference redshifts $z=3$ and $z=4.3$
for which many measurements have been performed. 

\item{b)} The clumping factor $f$ enters in a linear way in eq. (3.1.18):
thus, the presence of some inhomogeneities
($f > 1$) in the intergalactic medium increases the GP optical depth; 
a clumped configuration of the IGM makes it 
more opaque to the radiation.

\item{c)} I tested three values of the IGM density parameter, i.e.
$\Omega_{IGM}=0.010$,~$0.015,~0.020$; values larger than 0.020 produce
a too high optical depth $\tau_{GP}$, disagreeing with the results
of Giallongo et al. [39] (that however are affected by significant 
uncertainties). The disagreement is particularly serious in the 
inhomogeneous case ($f > 1$) with a medium / high absorption.

I can obtain a satisfactory agreement with the data of Giallongo et al. [39]
for $\Omega_{IGM}=0.010$ and a low absorption level: smaller values 
of $\Omega_{IGM}$ are in principle acceptable but, as I will show,
they do not allow to obtain a consistent value of the diffuse medium density 
parameter, $\Omega_{D}=\Omega_{IGM}-\Omega_{Ly\alpha}$.

\vskip 7mm
\centerline {\bf 4 CHARACTERISTICS OF THE $\lia$ CLOUDS}
\vskip 7mm
\centerline {\bf 4.1 THE THEORY}
\vskip 7mm
Here I want to discuss the characteristics of the IGM regions presenting very 
large overdensities, $\rho /\overline \rho \geq 10$, i.e. the configurations 
known as $\lia$ clouds; many informations can be obtained by studying 
the $\lia$ absorption phenomena within these regions. 
A very interesting idea is the possibility that these clouds are expanding 
and adiabatically cooled: the observations of Pettini et al. [14]
suggested that the $\lia$ forest lines with a low 
column density $N_{HI}$ ($N_{HI}\leq 10^{14}~cm^{-2}$)
have also very small velocity widths $b$.
They found a positive correlation between $b$ and $N_{HI}$; typical values 
for these parameters are [14], [15]:
$$
< b >~\sim ~11~\pm~3~km~sec^{-1}~~~~~~~~~~~~~~for ~~~~log~N_{HI}~=~13;
\eqno(4.1.1)
$$
$N_{HI}$ is expressed in $cm^{-2}$
and $b$ (also called `` Doppler parameter'') is expressed as
$$
b^{2}~=~V^{2}_{bulk}~+~{2kT\over m},\eqno(4.1.2)
$$
i.e. it is connected to the cloud characteristics, as the temperature 
and the bulk velocity.

Generally, these clouds are considered as small, dense and very filamentous 
structures, rather than large and highly ionized objects;
however, a different interpretation has been proposed in [15], [40]
where one assumes that these low density
clouds are indeed expanding structures, which progressively cool in an adiabatic
way; the expansion cooling may bring highly ionized clouds out of the
thermal ionization equilibrium, thus enabling lower values of temperature
($T << 10^{4}~K^{\circ}$) and of the velocity widths but not 
necessarily a low ionization level.

In the following, I will investigate the characteristics of $\lia$ clouds
by studying their absorption of the PBH photons; in particular, I will
calculate the ionization efficiency, the HI column density and the 
baryonic content of the $\lia$ clouds, expressed by the
density parameter $\Omega_{Ly\alpha}$.
We will see that, in such a model of PBH-induced reionization, 
senseful results for $\Omega_{Ly\alpha}$ can be obtained 
only in presence of expanding and adiabatically cooled clouds; thus,
my results seem to support the hypotheses of ref. [15].
\vskip 7mm
\centerline {\bf 4.2 IONIZATION EFFICIENCY, HI DENSITY}

\centerline {\bf AND DENSITY PARAMETER $\Omega_{Ly\alpha}$}
\vskip 7mm
Here I will recall the main equations that are useful to study the 
ionization problem.
In the case of photoionization equilibrium between the 
cloud gas and the ionizing radiation and for a ionizing flux given 
by eq. (3.1.15), the ionization efficiency is [32]:
$$
G_{H}~=~\int^{\infty}_{\nul}~{4\pi J_{\nu}\sigma_{\nu}\over h_{P}}
~{d\nu\over\nu}~{1\over J_{912}},\eqno(4.2.1)
$$
where $h_{P}$ is the Planck constant, $J_{\nu}/ J_{912}$ is given by 
eq. (3.1.19) and the hydrogen photoionization cross section is [41]
$$
\sigma_{\nu}~=~A_{0}~\Big(~{\nul\over\nu}~\Big)^{4}
~{exp~[4-((4\arctan\varepsilon) /
\varepsilon )]\over {1-exp~(-2\pi / \varepsilon)}},\eqno(4.2.2)
$$
holding for $\nu = \nul$ and with
$$
A_{0}~=~{2^{8}\pi\over 3 e^{4}}~({1\over 137})~\pi r_{0}^{2}~=~6.30~\times 
10^{-18}~cm^{2},\eqno(4.2.3)
$$
$$
\varepsilon~=~\sqrt{{\nu\over\nul}~-~1}.\eqno(4.2.4)
$$
Eq. (4.2.1) can be recast in the simpler form
$$
G_{H}~\sim~2.11\times 10^{6}~\nul~\int^{\numa}_{\nul}~{\sigma_{\nu}\over
\nu^{2}}~d\nu ,\eqno(4.2.5)
$$
where a suitable upper cut in the integration
($\numa = 1.\times 10^{17}~sec^{-1}$) is inserted in order to numerically 
handle the integration, after testing that this choice does not seriously 
influence the final result.
I obtain:
$$
G_{H}~=~3.52\times 10^{-12}~erg^{-1}~sec.\eqno(4.2.6)
$$
The total hydrogen gas density $n_{H,c}$ of a spherical cloud of average 
column density $N_{HI}=10^{14}~cm^{-2}$ is related to its diameter $D$
in the following way [32]:
$$
n_{H,c}(z)~=~\Bigg [~{3~G_{H}J_{912}(z)(10^{14}~cm^{-2})\over 
2 ~(1+2\chi)~\alpha_{A}(T_{c})~D}~\Bigg ]^{1/2},\eqno(4.2.7)
$$
where $\alpha_{A}= 4.2\times 10^{-13}~T_{4,c}^{-0.75}~cm^{3}~sec^{-1}$
is the coefficient of recombination to all the levels of hydrogen [41],
$T_{4, c}$ is the cloud temperature $(K)$ divided by $10^{4}$ and
$\chi$ is the ratio $He/H$, equal to $1/12$.
The column density $N_{HI}$, averaged over all the lines of sight, is 
connected to the diameter $D$ of the cloud through the following formula [32]:
$$
<N_{HI}>~=~{2\over 3}~n_{H,c}~D,\eqno(4.2.8)
$$
and the $\lia$ clouds density parameter, $\Omega_{Ly\alpha}$ is
$$
\Omega_{Ly\alpha}~=~f_{c}~{n_{H,c}\over n_{H,crit}};\eqno(4.2.9)
$$
here the closure hydrogen density is
$$
n_{H,crit}(0)~=~{3H_{0}^{2}\over 8\pi G m_{H} (1+4\chi)}~=~
2.11\times 10^{-6}~h^{2}~cm^{-3},\eqno(4.2.10)
$$
and the volume filling factor $f_{c}$ of the $\lia$ forest is [32]:
$$
f_{c}(z)~\sim~{2H_{0}\over c}~D(z)~(1+z)^{4.4}~(1+2q_{0}z)^{0.5}.\eqno(4.2.11)
$$
Finally, the size of the $\lia$ clouds is connected to the Doppler
parameter $b$ by the approximate formula [32]:
$$
D(z)~\sim~<b>~[H_{0}(1+z)~(1+2q_{0}z)^{0.5}]^{-1}.\eqno(4.2.12)
$$
Choosing the standard values $T_{c}=2\times 10^{4}~K^{\circ}$ and 
$<b>=35~km~sec^{-1}$ [15],
I obtained for the cloud hydrogen gas density $n_{H,c}$ and 
for the column density $N_{HI}$ the values listed in tab. 6a 
for two reference redshifts, $z=2.5$ and $z=0.0$.  

The baryonic density parameter at the present time 
for the $\lia$ clouds I calculated from eq. (4.2.9) is 
$$
\Omega_{Ly\alpha}~=~0.016;\eqno(4.2.13)
$$
This high value means that a large fraction 
of baryons resides in the $\lia$ clouds, while the diffuse component
of the IGM ($\Omega_{D}~=~\Omega_{IGM}~-~\Omega_{Ly\alpha}$) should have 
a density parameter totally negligible. A more reliable result can be 
obtained if we consider the clouds as dynamically expanding structures,
a possibility that also could explain the (probably) observed
correlation between the Doppler parameter $b$ and the column density $N_{HI}$.

In this case, the temperature $T_{c}$ lowers to $200~K^{\circ}$ and the size 
grows up to 1 Mpc; the results I obtain are listed in tab. 
6b. For this choice of $T_{c}$ and $D$, the present density parameter is 
$$
\Omega_{Ly\alpha}~=~0.010,\eqno(4.2.14)
$$
in all the cases of low, medium and high absorption (see tab. 7).
Consequently, the diffuse medium should have a present density parameter 
$$
0~\leq~\Omega_{D}~\leq 0.010,\eqno(4.2.15)
$$
depending on the effective value chosen for $\Omega_{IGM}$.

Above, I calculated the hydrogen density for the clumped component of the IGM;
now, the Gunn Peterson optical depth $\tau_{GP}(z)$ theoretically obtained 
in the previous section, can be used in eq. (3.1.17) in order to predict 
the density $n_{HI,~d}$ of its diffuse component.

Eq. (3.1.17) can be rewritten as follows: 
$$
\tau_{GP}~=~{8.28\times 10^{10}~n_{HI,~d}(z)\over (1+z)^{3/2}},\eqno(4.2.16)
$$
and therefore:
$$
n_{HI,~d}(z)~=~1.21\times 10^{-11}~\tau_{GP}(z)~(1+z)^{3/2}.\eqno(4.2.17)
$$
The density $n_{HI,~d}$ calculated for three reference redshifts,
$z=0,~z=2$ and $z=2.64$ are listed in tabs. 8a, 8b and 8c for 
$\Omega_{IGM}=0.010, ~0.015, ~0.020$ and for the cases of low, medium 
and high absorption.

\vfill\eject

\centerline {\bf 5. THE HE II GUNN PETERSON TEST}
\vskip 7mm
An interesting extension of the Gunn Peterson test is the study of 
the HeII $304~\AA$ resonance line of the singly ionized helium; both 
the theoretical and the experimental efforts are converging on this problem
since the appearance of the results of Jakobsen et al. [16], concerning the
probable detection of the HeII~$304~\AA$ Gunn Peterson absorption from the
IGM at $z\sim 3.2$. 

The primordial helium in the intergalactic medium should be strongly 
ionized at high redshifts and it should be mainly seen in the form of 
singly ionized HeII, rather than neutral HeI [16];
this fact has been confirmed by the failure in detecting 
the Gunn Peterson absorption in the HeI $584~\AA$ line of the neutral 
helium [42].

The HeII $\lia$ resonance line is redshifted at 
$\lambda_{obs}\sim 304~(1+z)~\AA$ and therefore it is observable in the extreme
UV; thus, the He~II Gunn Peterson test could potentially give a lot of 
informations on the spectrum of the UV background at very short wavelengths.
Distant quasars at redshifts $z>3$ are the most useful objects to
study, in order to detect this effect in an experimentally suitable range of 
wavelengths, $\lambda\geq 1200~\AA$, where the absorption by hydrogen 
in our galaxy does not produce noise.

From a theoretical point of view, the same kind of analysis used for the 
hydrogen can be performed for HeII, with only few differences;
firstly, the HeII ionization edge is located at $\lambda=228~\AA$,
corresponding to a frequency $\nu_{L,~He}=1.31\times 10^{16}~sec$; from eqs.
(3.1.17) and (3.1.18) it is possible to estimate [32]
the ratios of the intensities of the
diffuse radiation  field at the hydrogen Lyman edge ($\nu_{L, ~H}= c/ 912~\AA$)
and at the helium edge ($\nu_{L, ~He}= c/ 228~\AA$); for a low attenuation
model, Meiksin and Madau found [32]:
$$
J_{912} ~/ ~J_{228}~\sim ~8,\eqno(5.1)
$$
while in a medium attenuation model [28] one approximately has
$$
J_{912} ~/ ~J_{228}~\sim ~25.\eqno(5.2)
$$
From eqs. (5.1) and (5.2) I simply obtained the intensity $J_{228}$,
that I plotted in fig. 7.

Using these intensities in eq. (3.1.18) I obtain the results shown 
in tabs. 9a, 9b, 10a, 10b and 11a, 11b for the Gunn Peterson optical 
depth at two reference redshifts, $z=3$ and $z=4.3$ and for 
$\Omega_{IGM}=0.010,~0.015, ~0.020$. A plot of $\tau_{GP}(z)$ 
for the values of $\Omega_{IGM}$ and both in the homogeneous and 
inhomogeneous cases is shown in figs. 8a, 8b, 9a, 9b, 10a and 10b;
finally, in figs. 11a and 11b I resumed the behaviour of
$\tau_{GP}(z)$ for all the values of $\Omega_{IGM}$ and for a 
fixed attenuation level.
\vskip 5mm
The analysis of \S 4.2 can be repeated, simply observing 
that the $\lia$ clouds absorption process is just modified as follows:
the He absorption cross section is connected to the H cross 
section through the relation [8]:
$$
\sigma_{HeII}(\nu)~=~{1\over 4}~\sigma_{H}~\big( {\nu\over 4}\big);
\eqno(5.3)
$$
The ionization efficiency $G_{He}$ can be simply calculated by using
eq. (5.3) in eq. (4.2.1) and changing the intensities: thus, I calculate
$$
G_{He}~=~2.25\times 10^{-10}~erg^{-1}~sec,\eqno(5.4)
$$
Finally, the helium density $n_{He,~c}$ and the column density $N_{HeII}$
are given by some equations analogous to eqs. (4.2.7), (4.2.8),
with a ionization efficiency given by eq. (5.4), an intensity $J_{228}$
coming from eqs. (5.1) (LA case) and (5.2) (MA case) and a coefficient
of recombination for He given by [41]:
$$
\alpha_{A}(T)~=~2.607~\times ~10^{-13}~T_{4}^{-0.75}.\eqno(5.5)
$$
The results are listed in tab. 12a for $T_{c}=2\times 10^{4}~K^{\circ}$
and in tab. 12b for adiabatically expanding, cooled 
clouds at $T_{c}=200~K^{\circ}$.

A comparison of the Gunn Peterson optical
depth here obtained with the experimental value of Jakobsen et al. [16]
($\tau_{HeII} > 1.7$ at $z\sim 3$) is also possible but this lower limit
should be considered with some prudence: indeed, it is 
very conservative and probably affected by some uncertainties, due to the 
fact that the Authors cannot distinguish between HeII 
absorption coming from the line blanketing in the discrete lines of the
$\lia$ forest and the real Gunn Peterson trough.

Looking at tabs. 9a, 9b, 10a, 10b and 11a, 11b we can remark that 
in such a PBH-induced model of reionization, the Gunn Peterson
optical depths obtained for HeII are quite similar to the ones for HI:
in the case of inhomogeneous clouds with a medium absorption level 
and for $\Omega_{IGM}=0.020$, the maximum value for $\tau_{GP~HeII}$ at 
$z=3$ is 0.76.
The disagreement with the lower limit of ref. [16] should be
tested in the light of future, more precise measurements. 
\vskip 7mm
\centerline {\bf {6. CONCLUSIONS}}
\vskip 7mm
In this paper I assumed that the quantum evaporation of Primordial
Black Holes having a mass $M\sim 10^{14}~g$ provides 
the radiation that photoionizes the intergalactic medium at some early epochs
in the past.

This particular model has been tested in the light of 
the Gunn Peterson test for both HI and HeII: I calculated the
optical depths associated with the resonant $\lia$
absorption by clouds and I compared my theoretical results with the last
experimental data. 
The agreement is satisfactory, particularly if we assume an inhomogeneous 
IGM with a density parameter $\Omega=0.010$ and a low attenuation 
by the clouds.
In any case, a constraint on $\Omega_{IGM}$ can be inferred from this 
analysis: the agreement 
theory / experiment is possible only if $\Omega_{IGM} <0.020$.

I also studied the characteristics of the $\lia$ clouds and I found 
that my model confirms the possibility they are expanding and adiabatically
cooled.

A comparison theory / experiment for the HeII Gunn Peterson effect
is certainly premature, due to the paucity of the experimental results
regarding HeII: the HeII optical depth I calculated is 
smaller than the lower limit claimed by Jakobsen et al. [16]
($\tau_{GP} >1.7$) and near to the values I found for HI. Finally,
I can conclude that the reionization model here proposed seems to work quite 
well; anyway, a better knowledge of the ultimate fate of the PBHs relics,
particularly of their final masses, surely represents a definitive
test for it.
\vskip 7mm
\centerline{{\bf ACKNOWLEDGEMENTS}}
\vskip 7mm
This work has been financially supported by the University of Milano
that I want to thank; in particular, a sincere and affectionate
thank goes to all the people in the Astrophysics Section of the University 
of Milano for their friendly support and sympathy.
I thank also the Queen Mary and Westfield College for its hospitality.

\vskip 1cm
\centerline {\bf{REFERENCES}}
\vskip 8mm

\item{[1]} Gunn, J.E., Peterson, B.A., Astroph. J., 1965, {\bf 142}, 1633.

\item{[2]} Schmidt, M., Astroph. Journ., 1965, {\bf 141}, 1295.

\item{[3]} Davidson, A., Hartig, G.F., Fastie, W.G., Nature, 1977, {\bf 269}, 
203.

\item{[4]} Steidel, C.C., Sargent, W.L.W., Astroph. Journ., 1987, {\bf 318}, 
L11.

\item{[5]} Ostriker, J.P., Ikeuchi, S., Astroph. Journ., 1983, {\bf 268}, 
L63.

\item{[6]} Ikeuchi, S., Ostriker, J.P., Astroph. Journ., 1986, {\bf 301}, 
522.

\item{[7]} Chiang, W., Ryu, D., Vishniac, E.T., Astroph. Journ., 1989, 
{\bf 339}, 603.

\item{[8]} Miralda-Escud\'e, J., Ostriker, J.P., Astroph. Journ., 1990, 
{\bf 350}, 1.

\item{[9]} Carswell, R.F., Morton, D.C., Smith, M.G., Stockton, A. N.,
Turnshek, D.A., Weymann, R.J., Astroph. Journ., 1984, {\bf 278}, 486;

\item{} Tytler, D., Astroph. Journ., 1987, {\bf 321}, 69;

\item{} Murdoch, H.,S., Hunstead, R.W., Pettini, M., Blades, J.C., 
Astroph. Journ., 1986, {\bf 309}, 19.

\item{[10]} Bajtlik, S., Duncan, R.C., Ostriker, J.P., Astroph. Journ., 
1988, {\bf 327}, 570.

\item{[11]} Gibilisco, M., Intern. Journ. Of Mod. Phys. 1995, {\bf 10A}, 3605.

\item{[12]} Gibilisco, M., ''Reionization of the Universe induced by 
Primordial Black Holes '', Accept. for Pub. in Intern. Journ. of Mod. Phys A,
May 1996.
                           
\item{[13]} Gibilisco, M., ''The influence of quarks and gluons jets coming
from Primordial Black Holes on the Reionization of the Universe'',
submitted to Annals of Physics, 1996b.

\item{[14]} Pettini, M., Hunstead, R. W., Smith L.J., Mar, D.P., 
MNRAS, 1990, {\bf 246}, 545.

\item{[15]} Duncan, R.C., Vishniac, E.T., Ostriker, J.P.,
Astroph. Journ., 1991, {\bf 368}, L1.

\item{[16]} Jakobsen, P., Boksenberg, A., Deharveng, J.M., Greenfield, P.,
Jedrzejewski, R., Paresce, F., Nature, 1994, {\bf 370}, 35.

\item {[17]} Novikov, I., ''Black Holes and the Universe'', 1990, 
Cambridge Univ. Press.

\item{[18]} Hawking, S.W., Commun. Math. Phys., 1975, {\bf 43}, 199.

\item{[19]} Carr, B.J., Astroph. J., 1976, {\bf 206}, 8.

\item{[20]} Mac Gibbon, J.H., Carr, B.J., Astroph. J., 1991, {\bf 371}, 447.

\item{[21]} Mac Gibbon, J.H., Webber, B.R., Phys. Rev., 1990, {\bf D41}, 3052.

\item{[22]} Mac Gibbon, J. H., Phys. Rev., 1991, {\bf D44}, 376.

\item{[23]} Mather, J.C. et al., Astroph. J., 1994, {\bf 420}, 439.

\item{[24]} Page, D. N., Phys. Rev., 1976, {\bf D13}, 198.

\item{[25]} Page, D. N., Phys. Rev., 1977, {\bf D16}, 2402.

\item{[26]} Carr, B. J., Astronomical and Astroph. Transactions, 1994, 
Vol. {\bf 5}, 43.

\item{[27]} Bechtold, J., Weymann R. J., Lin, Z., Malkan, M.,
Astroph. J., 1987, {\bf 315}, 180.

\item{[28]} Madau, P., Astroph. J., 1992, {\bf 389}, L1.

\item{[29]} Songaila, A., Cowie, L. L., Lilly, S. J., 
Astroph. J., 1990, {\bf 348}, 371.

\item{[30]} Miralda-Escud\'e, J., Ostriker, J. P., Astroph. J., 1992,
{\bf 392}, 15.

\item{[31]} Shapiro, P.R., Giroux, M.L., Astroph. J., 1987, {\bf 321}, L107.

\item{[32]} Meiksin, A., Madau, P., Astroph. J., 1993, {\bf 412}, 34.

\item{[33]} Misner, C. W., Thorne, K.S., Wheeler, J. A.: ''Gravitation'', 1973,
p. 738, W. H. Freeman and Co., San Francisco.

\item{[34]} Paresce, F., McKee, C., Bowyers, S., Astroph. J., 1980, {\bf 240}, 
387.

\item{[35]} Haehnelt, M.G., Preprint Babbage astro-ph/9512024, December 
1995.
                        
\item{[36]} Cen, R., Miralda-Escud\'e, J., Ostriker, J. P., Rauch, M.,
Astroph. J., 1994, {\bf 437}, L9.

\item{[37]} Foltz, C.B., Weymann, R.J., R\"oser, H.J., Chaffee, F.H.,
Astroph. J., 1984, {\bf 281}, L1.

\item{[38]} Carlberg, R.G., Couchman, H.M.P., Astroph. J., 1989, {\bf 340}, 47.
                        
\item{[39]} Giallongo, E. et al., Astroph. J., 1994, {\bf 425}, L1.

\item{[40]} Bond, J.R., Szalay, A.S., Silk, J., Astroph. J., 1988,
{\bf 324}, 627.

\item{[41]} Osterbrock, D.E., ''Astrophysics of Gaseous Nebulae and AGN'',1989,
Oxford Univ. Press.

\item{[42]} Beaver, E.A., Astroph. J., 1991, {\bf 337}, L1;

\item{} Tripp, T.M., Green, R.F., Bechtold, J., Astroph. J., 1990, {\bf 364}, 
L29;

\item{} Reimers, D. et al., Nature, 1992, {360}, 561.

\vfill\eject
{\bf Tab. 1: Clumping function f coming from a numerical simulation of
Carlberg and Couchman (Carlberg \& Couchman 1989)}
\vskip 0.5cm
{\offinterlineskip
\tabskip=0pt
\halign{ \strut
	 \vrule#&
\quad    \bf# 
              \hfil \quad    &
	 \vrule#&
\quad	 \hfil #  &
	 \vrule#
	 \cr
\noalign{\hrule}
&~~~~~~~~~&&~~~~~~~~~~~ ~     ~~~~~&\cr
& f       &&~   z                  &\cr
&~~~~~~   &&~~~~~~~~~~~~           &\cr
\noalign {\hrule}
&	  &&                       &\cr
&  7.42   &&   0.00                &\cr
&	  &&                       &\cr
&  7.98   &&   0.81                &\cr
&	  &&                       &\cr
&  6.93   &&   1.36                &\cr
&	  &&                       &\cr
&  2.48   &&   2.80                &\cr
&	  &&                       &\cr
&  1.74   &&   4.63                &\cr
&	  &&                       &\cr
\noalign{\hrule}
}}
\vfill\eject
{\bf Tab. 2a: Gunn Peterson optical depth: homogeneous case ($f=1$) with
a density parameter $\Omega_{IGM}=0.010$}

\vskip 0.5cm
{\offinterlineskip
\tabskip=0pt
\halign{ \strut
	 \vrule#&
\quad    \bf# 
              \hfil \quad    &
	 \vrule#&
\quad	 \hfil #  &
	 \vrule#&
\quad	 \hfil # &
	 \vrule#
	 \cr
\noalign{\hrule}
&~~~~~~~~~&&~~~~~~~~~~~ ~      &&~~~~~~~~~~                                 &\cr
& ABSORPTION MODEL   && $z=3$             && $z=4.3$                        &\cr
&~~~~~~              &&~~~~~~~~~~~~       &&~~~~~~~~~~~~~~~~~               &\cr
\noalign {\hrule}
&~~~~~~              &&~~~~~~~~~~~~       &&~~~~~~~~~~~~~~~~~               &\cr
& Low Abs.           && 0.002             && 0.014                          &\cr
&~~~~~~              &&~~~~~~~~~~~~       &&~~~~~~~~~~~~~~~~~               &\cr
& Medium Abs.        && 0.003             && 0.030                          &\cr
&~~~~~~              &&~~~~~~~~~~~~       &&~~~~~~~~~~~~~~~~~               &\cr
& High Abs.          && 0.003             && 0.036                          &\cr
&                    &&                   &&                                &\cr
\noalign{\hrule}
 }}
\vskip 1cm
{\bf Tab. 2b: Gunn Peterson optical depth: inhomogeneous case ($f>1$) with
a density parameter $\Omega_{IGM}=0.010$}

\vskip 0.5cm
{\offinterlineskip
\tabskip=0pt
\halign{ \strut
	 \vrule#&
\quad    \bf# 
              \hfil \quad    &
	 \vrule#&
\quad	 \hfil #  &
	 \vrule#&
\quad	 \hfil # &
	 \vrule#
	 \cr
\noalign{\hrule}
&~~~~~~~~~&&~~~~~~~~~~~ ~      &&~~~~~~~~~~                                 &\cr
& ABSORPTION MODEL   && $z=3$             && $z=4.3$                        &\cr
&~~~~~~              &&~~~~~~~~~~~~       &&~~~~~~~~~~~~~~~~~               &\cr
\noalign {\hrule}
&~~~~~~              &&~~~~~~~~~~~~       &&~~~~~~~~~~~~~~~~~               &\cr
& Low Abs.           && 0.005             && 0.026                          &\cr
&~~~~~~              &&~~~~~~~~~~~~       &&~~~~~~~~~~~~~~~~~               &\cr
& Medium Abs.        && 0.008             && 0.057                          &\cr
&~~~~~~              &&~~~~~~~~~~~~       &&~~~~~~~~~~~~~~~~~               &\cr
& High Abs.          && 0.008             && 0.068                          &\cr
&                    &&                   &&                                &\cr
\noalign{\hrule}
 }}
\vfill\eject
{\bf Tab. 3a: Gunn Peterson optical depth: homogeneous case 

($f=1$) with
a density parameter $\Omega_{IGM}=0.015$}

\vskip 0.5cm
{\offinterlineskip
\tabskip=0pt
\halign{ \strut
	 \vrule#&
\quad    \bf# 
              \hfil \quad    &
	 \vrule#&
\quad	 \hfil #  &
	 \vrule#&
\quad	 \hfil # &
	 \vrule#
	 \cr
\noalign{\hrule}
&~~~~~~~~~&&~~~~~~~~~~~ ~      &&~~~~~~~~~~                                 &\cr
& ABSORPTION MODEL   && $z=3$             && $z=4.3$                        &\cr
&~~~~~~              &&~~~~~~~~~~~~       &&~~~~~~~~~~~~~~~~~               &\cr
\noalign {\hrule}
&~~~~~~              &&~~~~~~~~~~~~       &&~~~~~~~~~~~~~~~~~               &\cr
& Low Abs.           && 0.004             && 0.031                          &\cr
&~~~~~~              &&~~~~~~~~~~~~       &&~~~~~~~~~~~~~~~~~               &\cr
& Medium Abs.        && 0.007             && 0.068                          &\cr
&~~~~~~              &&~~~~~~~~~~~~       &&~~~~~~~~~~~~~~~~~               &\cr
& High Abs.          && 0.007             && 0.081                          &\cr
&	             &&                   &&                                &\cr
  \noalign{\hrule}
 }}
\vskip 1cm
{\bf Tab. 3b: Gunn Peterson optical depth: inhomogeneous case ($f>1$) with
a density parameter $\Omega_{IGM}=0.015$}

\vskip 0.5cm
{\offinterlineskip
\tabskip=0pt
\halign{ \strut
	 \vrule#&
\quad    \bf# 
              \hfil \quad    &
	 \vrule#&
\quad	 \hfil #  &
	 \vrule#&
\quad	 \hfil # &
	 \vrule#
	 \cr
\noalign{\hrule}
&~~~~~~~~~&&~~~~~~~~~~~ ~      &&~~~~~~~~~~                                 &\cr
& ABSORPTION MODEL   && $z=3$             && $z=4.3$                        &\cr
&~~~~~~              &&~~~~~~~~~~~~       &&~~~~~~~~~~~~~~~~~               &\cr
\noalign {\hrule}
&~~~~~~              &&~~~~~~~~~~~~       &&~~~~~~~~~~~~~~~~~               &\cr
& Low Abs.           && 0.010             && 0.057                          &\cr
&~~~~~~              &&~~~~~~~~~~~~       &&~~~~~~~~~~~~~~~~~               &\cr
& Medium Abs.        && 0.017             && 0.128                          &\cr
&~~~~~~              &&~~~~~~~~~~~~       &&~~~~~~~~~~~~~~~~~               &\cr
& High Abs.          && 0.017             && 0.152                          &\cr
&                    &&                   &&                                &\cr
  \noalign{\hrule}
 }}
\vfill\eject
{\bf Tab. 4a: Gunn Peterson optical depth: homogeneous case 

($f=1$) with
a density parameter $\Omega_{IGM}=0.020$}
\vskip 0.5cm
{\offinterlineskip
\tabskip=0pt
\halign{ \strut
	 \vrule#&
\quad    \bf# 
              \hfil \quad    &
	 \vrule#&
\quad	 \hfil #  &
	 \vrule#&
\quad	 \hfil # &
	 \vrule#
	 \cr
\noalign{\hrule}
&~~~~~~~~~&&~~~~~~~~~~~ ~      &&~~~~~~~~~~                                 &\cr
& ABSORPTION MODEL   && $z=3$             && $z=4.3$                        &\cr
&~~~~~~              &&~~~~~~~~~~~~       &&~~~~~~~~~~~~~~~~~               &\cr
\noalign {\hrule}
&~~~~~~              &&~~~~~~~~~~~~       &&~~~~~~~~~~~~~~~~~               &\cr
& Low Abs.           && 0.008             && 0.055                          &\cr
&~~~~~~              &&~~~~~~~~~~~~       &&~~~~~~~~~~~~~~~~~               &\cr
& Medium Abs.        && 0.012             && 0.121                          &\cr
&~~~~~~              &&~~~~~~~~~~~~       &&~~~~~~~~~~~~~~~~~               &\cr
& High Abs.          && 0.013             && 0.144                          &\cr
&                    &&                   &&                                &\cr
  \noalign{\hrule}
 }}
\vskip 1cm
{\bf Tab. 4b: Gunn Peterson optical depth: inhomogeneous case ($f>1$) with
a density parameter $\Omega_{IGM}=0.020$}

\vskip 0.5cm
{\offinterlineskip
\tabskip=0pt
\halign{ \strut
	 \vrule#&
\quad    \bf# 
              \hfil \quad    &
	 \vrule#&
\quad	 \hfil #  &
	 \vrule#&
\quad	 \hfil # &
	 \vrule#
	 \cr
\noalign{\hrule}
&~~~~~~~~~&&~~~~~~~~~~~ ~      &&~~~~~~~~~~                                 &\cr
& ABSORPTION MODEL   && $z=3$             && $z=4.3$                        &\cr
&~~~~~~              &&~~~~~~~~~~~~       &&~~~~~~~~~~~~~~~~~               &\cr
\noalign {\hrule}
&~~~~~~              &&~~~~~~~~~~~~       &&~~~~~~~~~~~~~~~~~               &\cr
& Low Abs.           && 0.018             && 0.102                          &\cr
&~~~~~~              &&~~~~~~~~~~~~       &&~~~~~~~~~~~~~~~~~               &\cr
& Medium Abs.        && 0.030             && 0.230                          &\cr
&~~~~~~              &&~~~~~~~~~~~~       &&~~~~~~~~~~~~~~~~~               &\cr
& High Abs.          && 0.030             && 0.270                          &\cr
&                    &&                   &&                                &\cr
  \noalign{\hrule}
 }}
\vfill\eject
{{\bf Tab. 5: Gunn Peterson optical depth: experimental data}
\vskip 0.5cm
{\offinterlineskip
\tabskip=0pt
\halign{ \strut
	 \vrule#&
\quad    \bf# 
         \hfil \quad    &
	 \vrule#&
\quad	 \hfil #  &
	 \vrule#&
\quad	 \hfil # &
	 \vrule#&
\quad	 \hfil #  &
	 \vrule#
	 \cr
\noalign{\hrule}
&~~~~~~        &&~~~~~~~~~~~&&~~~~~~~~~~~~ &&~~~~~~~~~                    &\cr
& $\tau_{GP}$  && $z$       && year        && Work                        &\cr 
&~~~~~~        &&~~~~~~~~~~~&&~~~~~~~~~~~~ &&                             &\cr
\noalign {\hrule}
&                    &&       &&           &&                              &\cr
& $0.01$             && 3.0   && 1992      && Giallongo et al. $^{a}$      &\cr
&                    &&       &&           &&                              &\cr
& $0.01\pm 0.03$     && 3.0   && 1994      && Giallongo et al. $^{b}$      &\cr
&                    &&       &&           &&                              &\cr
& $0.02\pm 0.03$     && 4.3   && 1994      && Giallongo et al. $^{b}$      &\cr
&                    &&       &&           &&                              &\cr
& $< 0.05$           && 3.8   && 1992      && Webb et al. $^{c}$           &\cr
&                    &&       &&           &&                              &\cr
& $0.04$             && 4.0   && 1992      && Webb et al. $^{c}$           &\cr
&                    &&       &&           &&                              &\cr
& $0.04\pm 0.01$     && 4.1   && 1992      && Webb et al. $^{c}$           &\cr
&                    &&       &&           &&                              &\cr
& $<0.02\pm 0.03$    && 2.64  && 1994      && Steidel and Sargent $^{d}$   &\cr
&                    &&       &&           &&                              &\cr
  \noalign{\hrule}
 }}
\vskip 8mm

\item{a)} Giallongo et al., APJ 398, L12, 1992.

\item{b)} Giallongo et al., APJ 425, L1, 1994.

\item{c)} Webb et al., MNRAS 255, 319, 1992.

\item{d)} Steidel and Sargent, APJ, 318, L11, 1987.

\vfill\eject
{\bf Tab. 6a: $Ly_{\alpha}$ clouds characteristics: hydrogen density $n_{H c}$
and column density $N_{HI}$ at a reference redshift $z=2.5$ and for 
$T_{C}=2\times 10^{4}~K^{\circ}$}

\vskip 0.5cm
{\offinterlineskip
\tabskip=0pt
\halign{ \strut
	 \vrule#&
\quad    \bf# 
              \hfil \quad    &
	 \vrule#&
\quad	 \hfil #  &
	 \vrule#&
\quad	 \hfil # &
	 \vrule#
	 \cr
\noalign{\hrule}
&~~~~~~~~~           &&~~~~~~~~~~~ ~         &&~~~~~~~                      &\cr
& ABS. MODEL         && $n_{Hc}~cm^{-3}$     && $N_{HI}~cm^{-2}$            &\cr
&~~~~~~              &&~~~~~~~~~~~~          &&~~~~~~~~~~~~~~~~~            &\cr
\noalign {\hrule}
&~~~~~~              &&~~~~~~~~~~~~          &&~~~~~~~~~~~~~~~~~            &\cr
& Low Abs.           && $1.04\times 10^{-4}$ && $2.28\times 10^{19}$        &\cr
&~~~~~~              &&~~~~~~~~~~~~          && ~~~~~~~~~~~~~               &\cr
& Medium Abs.        && $8.56\times 10^{-5}$ && $1.88\times 10^{19}$        &\cr
&~~~~~~              &&~~~~~~~~~~~~          && ~~~~~~~~~~~~~               &\cr
& High Abs.          && $8.91\times 10^{-5}$ && $1.96\times 10^{19}$        &\cr
&~~~~~~              &&~~~~~~~~~~~~          && ~~~~~~~~~~~~~               &\cr
  \noalign{\hrule}
 }}
\vskip 1cm

{\bf Tab. 6b: $Ly_{\alpha}$ clouds characteristics: hydrogen density $n_{H c}$
and column density $N_{HI}$ at a reference redshift $z=2.5$ and for 
$T_{C}=200~K^{\circ}$}

\vskip 0.5cm
{\offinterlineskip
\tabskip=0pt
\halign{ \strut
	 \vrule#&
\quad    \bf# 
              \hfil \quad    &
	 \vrule#&
\quad	 \hfil #  &
	 \vrule#&
\quad	 \hfil # &
	 \vrule#
	 \cr
\noalign{\hrule}
&~~~~~~~~~           &&~~~~~~~~~~~ ~         &&~~~~~~~                      &\cr
& ABS. MODEL         && $n_{Hc}~cm^{-3}$     && $N_{HI}~cm^{-2}$            &\cr
&~~~~~~              &&~~~~~~~~~~~~          &&~~~~~~~~~~~~~~~~~            &\cr
\noalign {\hrule}
&~~~~~~              &&~~~~~~~~~~~~          &&~~~~~~~~~~~~~~~~~            &\cr
& Low Abs.           && $1.84\times 10^{-5}$ && $4.05\times 10^{18}$        &\cr
&~~~~~~              &&~~~~~~~~~~~~          && ~~~~~~~~~~~~~               &\cr
& Medium Abs.        && $1.52\times 10^{-5}$ && $3.35\times 10^{18}$        &\cr
&~~~~~~              &&~~~~~~~~~~~~          && ~~~~~~~~~~~~~               &\cr
& High Abs.          && $1.58\times 10^{-5}$ && $3.48\times 10^{18}$        &\cr
&~~~~~~              &&~~~~~~~~~~~~          && ~~~~~~~~~~~~~               &\cr
  \noalign{\hrule}
 }}

\vfill\eject
{\bf Tab. 7: Density parameter $\Omega_{Ly\alpha}$ for $Ly_{\alpha}$
clusters for a cluster temperature $T_{c}=200~K^{\circ}$}

\vskip 0.5cm
{\offinterlineskip
\tabskip=0pt
\halign{ \strut
	 \vrule#&
\quad    \bf# 
              \hfil \quad    &
	 \vrule#&
\quad	 \hfil #  &
	 \vrule#&
\quad	 \hfil # &
	 \vrule#
	 \cr
\noalign{\hrule}
&~~~~~~~~~&&~~~~~~~~~~~ ~      &&~~~~~~~~~~                                 &\cr
& ABS. MODEL         && $z=2.5$           && $z=0.0$                        &\cr
&~~~~~~              &&~~~~~~~~~~~~       &&~~~~~~~~~~~~~~~~~               &\cr
\noalign {\hrule}
&~~~~~~              &&~~~~~~~~~~~~       &&~~~~~~~~~~~~~~~~~               &\cr
& Low Abs.           && 0.013             && 0.010                          &\cr
&~~~~~~              &&~~~~~~~~~~~~       &&~~~~~~~~~~~~~~~~~               &\cr
& Medium Abs.        && 0.011             && 0.010                          &\cr
&~~~~~~              &&~~~~~~~~~~~~       &&~~~~~~~~~~~~~~~~~               &\cr
& High Abs.          && 0.012             && 0.010                          &\cr
&~~~~~~              &&~~~~~~~~~~~~       &&~~~~~~~~~~~~~~~~~               &\cr
  \noalign{\hrule}
 }}

\vfill\eject

{\bf Tab. 8a: Hydrogen density $n_{H,~d}$ at redshift $z=0$, $z=2.00$ and
$z=2.64$ characterizing the diffuse component of the 
IGM; $\Omega_{IGM}=0.010$ and $f>1$}
\vskip 0.5cm
{\offinterlineskip
\tabskip=0pt
\halign{ \strut
	 \vrule#&
\quad    \bf# 
              \hfil \quad    &
	 \vrule#&
\quad	 \hfil #  &
	 \vrule#&
\quad	 \hfil #  &
	 \vrule#&
\quad	 \hfil #  &
         \vrule#&
\quad	 \hfil #  &
	 \vrule#
	 \cr
\noalign{\hrule}
&~~~~~~~~~           &&~~~~~~~~~~~ ~           &&~~~~~~~                      
                     &&                                                     &\cr
& ABS. MODEL         && $n_{H,d}(z=0)~cm^{-3}$ && $n_{H,d}(z=2.00)~cm^{-3}$
                     && $n_{H,d}(z=2.64)~cm^{-3}$                           &\cr
&~~~~~~           &&~~~~~~~~~~~~             &&~~~~~~~~~~~~~~~~~            
                  &&                                                        &\cr
\noalign {\hrule}
&~~~~~~           &&~~~~~~~~~~~~             &&~~~~~~~~~~~~~~~~~            
                  &&                                                        &\cr
& Low Abs.        && $4.91\times 10^{-17}$   && $8.99\times 10^{-14}$       
                  && $2.47\times 10^{-13}$                                  &\cr
&~~~~~~           &&~~~~~~~~~~~~             &&~~~~~~~~~~~~~~~~~            
                  &&                                                        &\cr
& Medium Abs.     && $5.01\times 10^{-17}$   && $1.19\times 10^{-13}$       
                  && $3.83\times 10^{-13}$                                  &\cr
&~~~~~~           &&~~~~~~~~~~~~             &&~~~~~~~~~~~~~~~~~            
                  &&                                                        &\cr
& High Abs.       && $4.02\times 10^{-17}$   && $1.01\times 10^{-13}$       
                  && $3.62\times 10^{-13}$                                  &\cr
&~~~~~~           &&~~~~~~~~~~~~             &&~~~~~~~~~~~~~~~~~            
                  &&                                                        &\cr
  \noalign{\hrule}
 }}

\vfill\eject

{\bf Tab. 8b: Hydrogen density $n_{H,~d}$ at redshift $z=0$, $z=2.00$ and
$z=2.64$ characterizing the diffuse component of the 
IGM; $\Omega_{IGM}=0.015$ and $f>1$}
\vskip 0.5cm
{\offinterlineskip
\tabskip=0pt
\halign{ \strut
	 \vrule#&
\quad    \bf# 
              \hfil \quad    &
	 \vrule#&
\quad	 \hfil #  &
	 \vrule#&
\quad	 \hfil #  &
	 \vrule#&
\quad	 \hfil #  &
         \vrule#&
\quad	 \hfil #  &
	 \vrule#
	 \cr
\noalign{\hrule}
&~~~~~~~~~           &&~~~~~~~~~~~ ~           &&~~~~~~~                      
                     &&                                                     &\cr
& ABS. MODEL         && $n_{H,d}(z=0)~cm^{-3}$ && $n_{H,d}(z=2.00)~cm^{-3}$
                     && $n_{H,d}(z=2.64)~cm^{-3}$                           &\cr
&~~~~~~           &&~~~~~~~~~~~~             &&~~~~~~~~~~~~~~~~~            
                  &&                                                        &\cr
\noalign {\hrule}
&~~~~~~           &&~~~~~~~~~~~~             &&~~~~~~~~~~~~~~~~~            
                  &&                                                        &\cr
& Low Abs.        && $1.10\times 10^{-16}$   && $2.02\times 10^{-13}$       
                  && $5.71\times 10^{-13}$                                  &\cr
&~~~~~~           &&~~~~~~~~~~~~             &&~~~~~~~~~~~~~~~~~            
                  &&                                                        &\cr
& Medium Abs.     && $1.12\times 10^{-16}$   && $2.67\times 10^{-13}$       
                  && $8.62\times 10^{-13}$                                  &\cr
&~~~~~~           &&~~~~~~~~~~~~             &&~~~~~~~~~~~~~~~~~            
                  &&                                                        &\cr
& High Abs.       && $9.04\times 10^{-17}$   && $2.28\times 10^{-13}$       
                  && $8.15\times 10^{-13}$                                  &\cr
&~~~~~~           &&~~~~~~~~~~~~             &&~~~~~~~~~~~~~~~~~            
                  &&                                                        &\cr
  \noalign{\hrule}
 }}

\vfill\eject

{\bf Tab. 8c: Hydrogen density $n_{H,~d}$ at redshift $z=0$, $z=2.00$ and
$z=2.64$ characterizing the diffuse component of the 
IGM; $\Omega_{IGM}=0.020$ and $f>1$}
\vskip 0.5cm
{\offinterlineskip
\tabskip=0pt
\halign{ \strut
	 \vrule#&
\quad    \bf# 
              \hfil \quad    &
	 \vrule#&
\quad	 \hfil #  &
	 \vrule#&
\quad	 \hfil #  &
	 \vrule#&
\quad	 \hfil #  &
         \vrule#&
\quad	 \hfil #  &
	 \vrule#
	 \cr
\noalign{\hrule}
&~~~~~~~~~           &&~~~~~~~~~~~ ~           &&~~~~~~~                      
                     &&                                                     &\cr
& ABS. MODEL         && $n_{H,d}(z=0)~cm^{-3}$ && $n_{H,d}(z=2.00)~cm^{-3}$
                     && $n_{H,d}(z=2.64)~cm^{-3}$                           &\cr
&~~~~~~           &&~~~~~~~~~~~~             &&~~~~~~~~~~~~~~~~~            
                  &&                                                        &\cr
\noalign {\hrule}
&~~~~~~           &&~~~~~~~~~~~~             &&~~~~~~~~~~~~~~~~~            
                  &&                                                        &\cr
& Low Abs.        && $1.96\times 10^{-16}$   && $3.60\times 10^{-13}$       
                  && $1.01\times 10^{-12}$                                  &\cr
&~~~~~~           &&~~~~~~~~~~~~             &&~~~~~~~~~~~~~~~~~            
                  &&                                                        &\cr
& Medium Abs.     && $1.61\times 10^{-16}$   && $4.06\times 10^{-13}$       
                  && $1.45\times 10^{-12}$                                  &\cr
&~~~~~~           &&~~~~~~~~~~~~             &&~~~~~~~~~~~~~~~~~            
                  &&                                                        &\cr
& High Abs.       && $2.03\times 10^{-16}$   && $4.75\times 10^{-13}$       
                  && $1.53\times 10^{-12}$                                  &\cr
&~~~~~~           &&~~~~~~~~~~~~             &&~~~~~~~~~~~~~~~~~            
                  &&                                                        &\cr
  \noalign{\hrule}
 }}

\vfill\eject

{\bf Tab. 9a: HeII Gunn Peterson optical depth: homogeneous case ($f=1$) with
a density parameter $\Omega_{IGM}=0.010$}

\vskip 0.5cm
{\offinterlineskip
\tabskip=0pt
\halign{ \strut
	 \vrule#&
\quad    \bf# 
              \hfil \quad    &
	 \vrule#&
\quad	 \hfil #  &
	 \vrule#&
\quad	 \hfil # &
	 \vrule#
	 \cr
\noalign{\hrule}
&~~~~~~~~~&&~~~~~~~~~~~ ~      &&~~~~~~~~~~                                 &\cr
& ABSORPTION MODEL   && $z=3$             && $z=4.3$                        &\cr
&~~~~~~              &&~~~~~~~~~~~~       &&~~~~~~~~~~~~~~~~~               &\cr
\noalign {\hrule}
&~~~~~~              &&~~~~~~~~~~~~       &&~~~~~~~~~~~~~~~~~               &\cr
& Low Abs.           && 0.037             && 0.204                          &\cr
&~~~~~~              &&~~~~~~~~~~~~       &&~~~~~~~~~~~~~~~~~               &\cr
& Medium Abs.        && 0.190             && 1.420                          &\cr
&~~~~~~              &&~~~~~~~~~~~~       &&~~~~~~~~~~~~~~~~~               &\cr
\noalign{\hrule}
 }}
\vskip 1cm
{\bf Tab. 9b: HeII Gunn Peterson optical depth: inhomogeneous 
case ($f>1$) with
a density parameter $\Omega_{IGM}=0.010$}

\vskip 0.5cm
{\offinterlineskip
\tabskip=0pt
\halign{ \strut
	 \vrule#&
\quad    \bf# 
              \hfil \quad    &
	 \vrule#&
\quad	 \hfil #  &
	 \vrule#&
\quad	 \hfil # &
	 \vrule#
	 \cr
\noalign{\hrule}
&~~~~~~~~~&&~~~~~~~~~~~ ~      &&~~~~~~~~~~                                 &\cr
& ABSORPTION MODEL   && $z=3$             && $z=4.3$                        &\cr
&~~~~~~              &&~~~~~~~~~~~~       &&~~~~~~~~~~~~~~~~~               &\cr
\noalign {\hrule}
&~~~~~~              &&~~~~~~~~~~~~       &&~~~~~~~~~~~~~~~~~               &\cr
& Low Abs.           && 0.015             && 0.109                          &\cr
&~~~~~~              &&~~~~~~~~~~~~       &&~~~~~~~~~~~~~~~~~               &\cr
& Medium Abs.        && 0.079             && 0.758                          &\cr
&                    &&                   &&                                &\cr
\noalign{\hrule}
 }}
\vfill\eject
{\bf Tab. 10a: HeII Gunn Peterson optical depth: homogeneous case ($f=1$) with
a density parameter $\Omega_{IGM}=0.015$}

\vskip 0.5cm
{\offinterlineskip
\tabskip=0pt
\halign{ \strut
	 \vrule#&
\quad    \bf# 
              \hfil \quad    &
	 \vrule#&
\quad	 \hfil #  &
	 \vrule#&
\quad	 \hfil # &
	 \vrule#
	 \cr
\noalign{\hrule}
&~~~~~~~~~&&~~~~~~~~~~~ ~      &&~~~~~~~~~~                                 &\cr
& ABSORPTION MODEL   && $z=3$             && $z=4.3$                        &\cr
&~~~~~~              &&~~~~~~~~~~~~       &&~~~~~~~~~~~~~~~~~               &\cr
\noalign {\hrule}
&~~~~~~              &&~~~~~~~~~~~~       &&~~~~~~~~~~~~~~~~~               &\cr
& Low Abs.           && 0.083             && 0.460                          &\cr
&~~~~~~              &&~~~~~~~~~~~~       &&~~~~~~~~~~~~~~~~~               &\cr
& Medium Abs.        && 0.428             && 3.195                          &\cr
&~~~~~~              &&~~~~~~~~~~~~       &&~~~~~~~~~~~~~~~~~               &\cr
\noalign{\hrule}
 }}
\vskip 1cm
{\bf Tab. 10b: HeII Gunn Peterson optical depth: inhomogeneous 
case ($f>1$) with
a density parameter $\Omega_{IGM}=0.015$}

\vskip 0.5cm
{\offinterlineskip
\tabskip=0pt
\halign{ \strut
	 \vrule#&
\quad    \bf# 
              \hfil \quad    &
	 \vrule#&
\quad	 \hfil #  &
	 \vrule#&
\quad	 \hfil # &
	 \vrule#
	 \cr
\noalign{\hrule}
&~~~~~~~~~&&~~~~~~~~~~~ ~      &&~~~~~~~~~~                                 &\cr
& ABSORPTION MODEL   && $z=3$             && $z=4.3$                        &\cr
&~~~~~~              &&~~~~~~~~~~~~       &&~~~~~~~~~~~~~~~~~               &\cr
\noalign {\hrule}
&~~~~~~              &&~~~~~~~~~~~~       &&~~~~~~~~~~~~~~~~~               &\cr
& Low Abs.           && 0.035             && 0.246                          &\cr
&~~~~~~              &&~~~~~~~~~~~~       &&~~~~~~~~~~~~~~~~~               &\cr
& Medium Abs.        && 0.178             && 1.706                          &\cr
&                    &&                   &&                                &\cr
\noalign{\hrule}
 }}
\vfill\eject
{\bf Tab. 11a: HeII Gunn Peterson optical depth: homogeneous case ($f=1$) with
a density parameter $\Omega_{IGM}=0.020$}

\vskip 0.5cm
{\offinterlineskip
\tabskip=0pt
\halign{ \strut
	 \vrule#&
\quad    \bf# 
              \hfil \quad    &
	 \vrule#&
\quad	 \hfil #  &
	 \vrule#&
\quad	 \hfil # &
	 \vrule#
	 \cr
\noalign{\hrule}
&~~~~~~~~~&&~~~~~~~~~~~ ~      &&~~~~~~~~~~                                 &\cr
& ABSORPTION MODEL   && $z=3$             && $z=4.3$                        &\cr
&~~~~~~              &&~~~~~~~~~~~~       &&~~~~~~~~~~~~~~~~~               &\cr
\noalign {\hrule}
&~~~~~~              &&~~~~~~~~~~~~       &&~~~~~~~~~~~~~~~~~               &\cr
& Low Abs.           && 0.148             && 0.818                          &\cr
&~~~~~~              &&~~~~~~~~~~~~       &&~~~~~~~~~~~~~~~~~               &\cr
& Medium Abs.        && 0.760             && 5.679                          &\cr
&~~~~~~              &&~~~~~~~~~~~~       &&~~~~~~~~~~~~~~~~~               &\cr
\noalign{\hrule}
 }}
\vskip 1cm
{\bf Tab. 11b: HeII Gunn Peterson optical depth: inhomogeneous 
case ($f>1$) with
a density parameter $\Omega_{IGM}=0.020$}

\vskip 0.5cm
{\offinterlineskip
\tabskip=0pt
\halign{ \strut
	 \vrule#&
\quad    \bf# 
              \hfil \quad    &
	 \vrule#&
\quad	 \hfil #  &
	 \vrule#&
\quad	 \hfil # &
	 \vrule#
	 \cr
\noalign{\hrule}
&~~~~~~~~~&&~~~~~~~~~~~ ~      &&~~~~~~~~~~                                 &\cr
& ABSORPTION MODEL   && $z=3$             && $z=4.3$                        &\cr
&~~~~~~              &&~~~~~~~~~~~~       &&~~~~~~~~~~~~~~~~~               &\cr
\noalign {\hrule}
&~~~~~~              &&~~~~~~~~~~~~       &&~~~~~~~~~~~~~~~~~               &\cr
& Low Abs.           && 0.062             && 0.437                          &\cr
&~~~~~~              &&~~~~~~~~~~~~       &&~~~~~~~~~~~~~~~~~               &\cr
& Medium Abs.        && 0.317             && 3.032                          &\cr
&                    &&                   &&                                &\cr
\noalign{\hrule}
 }}
\vfill\eject
{\bf Tab. 12a: $Ly_{\alpha}$ clouds characteristics: helium density $n_{He c}$
and column density $N_{HeII}$ at a reference redshift $z=2.5$ and for 
$T_{C}=2\times 10^{4}~K^{\circ}$}

\vskip 0.5cm
{\offinterlineskip
\tabskip=0pt
\halign{ \strut
	 \vrule#&
\quad    \bf# 
              \hfil \quad    &
	 \vrule#&
\quad	 \hfil #  &
	 \vrule#&
\quad	 \hfil # &
	 \vrule#
	 \cr
\noalign{\hrule}
&~~~~~~~~~           &&~~~~~~~~~~~ ~         &&~~~~~~~                      &\cr
& ABSORPTION MODEL   && $n_{Hc}~cm^{-3}$     && $N_{HI}~cm^{-2}$            &\cr
&~~~~~~              &&~~~~~~~~~~~~          &&~~~~~~~~~~~~~~~~~            &\cr
\noalign {\hrule}
&~~~~~~              &&~~~~~~~~~~~~          &&~~~~~~~~~~~~~~~~~            &\cr
& Low Abs.           && $4.11\times 10^{-5}$ && $9.06\times 10^{18}$        &\cr
&~~~~~~              &&~~~~~~~~~~~~          && ~~~~~~~~~~~~~               &\cr
& Medium Abs.        && $1.92\times 10^{-5}$ && $4.23\times 10^{18}$        &\cr
&~~~~~~              &&~~~~~~~~~~~~          && ~~~~~~~~~~~~~               &\cr
  \noalign{\hrule}
 }}
\vskip 1cm

{\bf Tab. 12b: $Ly_{\alpha}$ clouds characteristics: helium density $n_{He c}$
and column density $N_{HeII}$ at a reference redshift $z=2.5$ and for 
$T_{C}=200~K^{\circ}$}

\vskip 0.5cm
{\offinterlineskip
\tabskip=0pt
\halign{ \strut
	 \vrule#&
\quad    \bf# 
              \hfil \quad    &
	 \vrule#&
\quad	 \hfil #  &
	 \vrule#&
\quad	 \hfil # &
	 \vrule#
	 \cr
\noalign{\hrule}
&~~~~~~~~~           &&~~~~~~~~~~~ ~         &&~~~~~~~                      &\cr
& ABSORPTION MODEL   && $n_{Hc}~cm^{-3}$     && $N_{HI}~cm^{-2}$            &\cr
&~~~~~~              &&~~~~~~~~~~~~          &&~~~~~~~~~~~~~~~~~            &\cr
\noalign {\hrule}
&~~~~~~              &&~~~~~~~~~~~~          &&~~~~~~~~~~~~~~~~~            &\cr
& Low Abs.           && $2.32\times 10^{-4}$ && $5.09\times 10^{19}$        &\cr
&~~~~~~              &&~~~~~~~~~~~~          && ~~~~~~~~~~~~~               &\cr
& Medium Abs.        && $1.08\times 10^{-4}$ && $2.38\times 10^{19}$        &\cr
&~~~~~~              &&~~~~~~~~~~~~          && ~~~~~~~~~~~~~               &\cr
  \noalign{\hrule}
 }}

\vfill\eject

{\bf FIGURE CAPTIONS:}

\vskip 5mm

\item {Fig. 1:} Average ionizing intensity per unit of frequency and steradiant
$J_{921}(z)$ at the hydrogen Lyman edge $\nu_{L}=c/912~\AA$ 
($erg~cm^{-3}~sec^{-1}~Hz^{-1}~sr^{-1}$; full line: low attenuation model;
dashed line: medium attenuation model; dotted line: high attenuation model.

\vskip 3mm

\item {Fig. 2:} Evolution of the temperature of the intergalactic medium
as a function of the redshift $z$ (eV).

\vskip 3mm

\item {Fig. 3a:} Gunn Peterson optical depth $\tau_{GP}(z)$ associated with the
resonant HI $\lia$ absorption for an inhomogeneous 
IGM with $\Omega_{IGM}=0.010$ and with a clumping factor $f <1$;
full line: low attenuation model;
dashed line: medium attenuation model; dotted line: high attenuation model.

\vskip 3mm

\item {Fig. 3b:} Gunn Peterson optical depth $\tau_{GP}(z)$ associated with the
resonant HI $\lia$ absorption for a homogeneous 
IGM with $\Omega_{IGM}=0.010$ and with a clumping factor $f =1$.

\vskip 3mm

\item {Figs. 4a-4b:} The same plots of figs. 3a and 3b but for 
$\Omega_{IGM}=0.015$.

\vskip 3mm

\item {Figs. 5a-5b:} The same plots of figs. 3a and 3b but for 
$\Omega_{IGM}=0.020$.

\vskip 3mm

\item {Fig. 6a:} Gunn Peterson optical depth associated with the
resonant HI $\lia$ absorption for the three values of $\Omega_{IGM}$
in the case of an inhomogeneous IGM ($f>1$) with low absorption;
full line: $\Omega_{IGM}=0.010$, dashed line: $\Omega_{IGM}=0.015$,
dotted line: $\Omega_{IGM}=0.020$.

\item {Fig. 6b:} The same plot of fig. 6a but in the case of medium 
attenuation.

\item {Fig. 6c:} The same plot of fig. 6a but in the case of high attenuation.

\vskip 3mm

\item {Fig. 7:} Average ionizing intensity per unit of frequency and steradiant
$J_{228}(z)$ at the Helium Lyman edge $\nu_{L}=c/228~\AA$ 
($erg~cm^{-3}~sec^{-1}~Hz^{-1}~sr^{-1}$; full line: low attenuation model;
dashed line: medium attenuation model; dotted line: high attenuation model.

\vskip 3mm

\item {Fig. 8a:} Gunn Peterson optical depth $\tau_{GP}(z)$ associated with the
resonant HeII $\lia$ absorption for an inhomogeneous 
IGM with $\Omega_{IGM}=0.010$ and with a clumping factor $f <1$;
full line: low attenuation model; dashed line: medium attenuation model.

\vskip 3mm

\item {Fig. 8b:} Gunn Peterson optical depth $\tau_{GP}(z)$ associated with the
resonant HeII $\lia$ absorption for a homogeneous 
IGM with $\Omega_{IGM}=0.010$ and with a clumping factor $f =1$.

\vskip 3mm

\item {Figs. 9a-9b:} The same plots of figs. 8a and 8b but for 
$\Omega_{IGM}=0.015$.

\vskip 3mm

\item {Figs. 10a-10b:} The same plots of figs. 8a and 8b but for 
$\Omega_{IGM}=0.020$.

\vskip 3mm

\item {Fig. 11a:} Gunn Peterson optical depth associated with the
resonant HeII $\lia$ absorption for the three chosen values of $\Omega_{IGM}$
in the case of an inhomogeneous IGM ($f>1$) with low absorption;
full line: $\Omega_{IGM}=0.010$, dashed line: $\Omega_{IGM}=0.015$,
dotted line: $\Omega_{IGM}=0.020$.

\item {Fig. 11b:} The same plot of fig. 6a but in 
the case of medium attenuation.

\bye